\documentclass[a4paper,11pt]{article}
\usepackage{pos}
\usepackage{physics}
\usepackage{feynmp}
\usepackage{multirow}
 
\makeatletter
\def\endfmffile{
  \fmfcmd{\p@rcent\space the end.^^J end.^^J endinput;}
  \if@fmfio
    \immediate\closeout\@outfmf
  \fi
  \ifnum\pdfshellescape>\z@
    \immediate\write18{mpost \thefmffile}
  \fi}
\makeatother

\title{Strange and charm contributions to the HVP from C$^{\star}$ boundary conditions}
\author[a]{Anian Altherr}
\author[b]{Lucius Bushnaq}
\author[c]{Isabel Campos}
\author[a]{Marco Catillo}
\author[d]{Alessandro Cotellucci}
\author[d,e,f,g]{Madeleine Dale}
\author[b]{Patrick Fritzsch}
\author[a]{Roman Gruber}
\author[a]{Javad Komijani}
\author[d,h]{Jens Lücke}
\author[a]{Marina Krstić Marinković}
\author[i]{Sofie Martins}
\author[d,h]{Agostino Patella}
\author[e,f]{Nazario Tantalo}
\author*[a]{Paola Tavella}

\affiliation[a]{Institut für Theoretische Physik, ETH Zürich, Wolfgang-Pauli-Str. 27, 8093 Zürich, Switzerland}
\affiliation[b]{School of Mathematics, Trinity College Dublin, Dublin 2, Ireland}
\affiliation[c]{Instituto de Física de Cantabria and IFCA-CSIC, Avda. de Los Castros s/n, 39005 Santander, Spain}
\affiliation[d]{Humboldt Universität zu Berlin, Institut für Physik and IRIS Adlershof, Zum Großen Windkanal 6, 12489
Berlin, German
}
\affiliation[e]{Università di Roma Tor Vergata, Dip. di Fisica, Via della Ricerca Scientifica 1, 00133 Rome, Italy}
\affiliation[f]{INFN, Sezione di Tor Vergata, Via della Ricerca Scientifica 1, 00133 Rome, Italy}
\affiliation[g]{University of Cyprus, Department of Physics, 1 Panepistimiou Street, 2109 Aglantzia, Nicosia, Cyprus}
\affiliation[h]{DESY, Platanenallee 6, D-15738 Zeuthen, German}
\affiliation[i]{CP3-Origins \& IMADA, University of Southern Denmark, Campusvej 55, 5230 Odense M, Denmark}

\emailAdd{ptavella@phys.ethz.ch}
\abstract{
{\centering
\includegraphics[width=0.12 \linewidth]{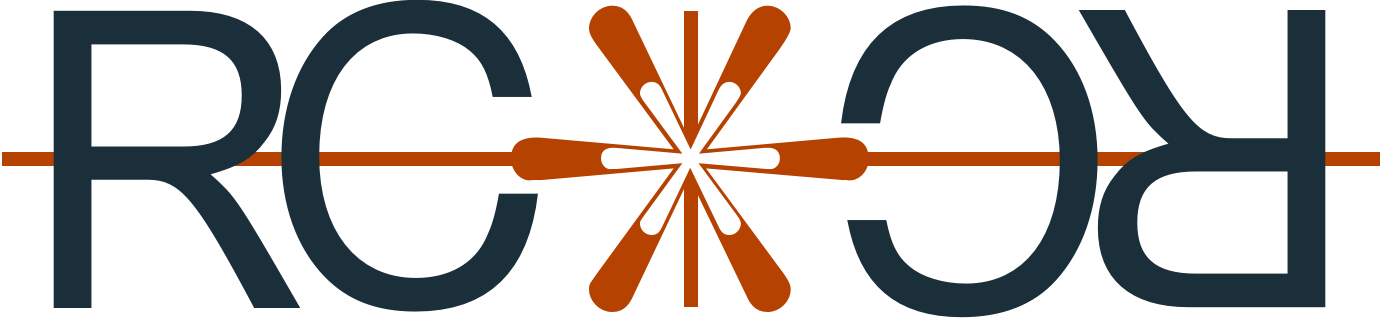}
\large{ collaboration} \par}
\vspace{10pt}

We present preliminary results for the determination of the leading strange and charm quark-connected contributions to the hadronic vacuum polarization contribution to the muon's $g-2$. Measurements are performed on the RC$^{\star}$ collaboration’s QCD ensembles, with $3+1$ flavors of $O(a)$ improved Wilson fermions and C$^{\star}$ boundary conditions. The HVP is computed on a single value of the lattice spacing and two lattice volumes at unphysical pion mass. In addition, we compare the signal-to-noise ratio for different lattice discretizations of the vector current.}

\FullConference{%
  The 39th International Symposium on Lattice Field Theory (Lattice2022),\\
  8-13th August, 2022 \\
  Bonn, Germany 
}

\begin{document}
\maketitle
\section{Introduction}
The anomalous magnetic moment of the muon is one of the quantities which is getting a
deal of attention in relation to new physics searches. The combined result of the BNL's E821 experiment \cite{PhysRevD.73.072003} and the first run of the E989 experiment at Fermilab \cite{PhysRevLett.126.141801} shows a precision of 0.35 ppm and a tension with the Standard Model's prediction \cite{Aoyama_2020} of 4.2 $\sigma$, if one does not include recent lattice determinations, most notably the result by the BMW collaboration \cite{Borsanyi_2021}. The next runs of the E989 experiment and the upcoming experiments at J-PARC \cite{https://doi.org/10.48550/arxiv.1901.03047} and CERN \cite{Abbiendi} aim to further reduce the experimental uncertainty. %experimental

Theoretically, the dominant source of uncertainty is the leading hadronic vacuum polarization. The most precise result for $a^{\mathrm{LO,HVP}}_{\mu}$ is obtained using the dispersive relations and the experimental data for the cross section of $e^+e^-$ to hadrons. Currently, the precision of the dispersive approach is about 0.6\% \cite{Aoyama_2020}. Independent results can be obtained using the lattice framework, which does not require experimental inputs and has started to produce competitive results for the muon's $g-2$. The most precise result from lattice simulations is the one from the BMW collaboration, which shows a precision of about 0.8\%  \cite{Borsanyi_2021}. The target precision on the HVP for the next few years is of few per mille. To achieve this precision, it is necessary to include the strong and electromagnetic isospin-breaking corrections, which contribute at the percent level. 

In this work, we present preliminary results for the leading connected contributions to HVP from strange and charm quarks. This is the first and necessary step for a long-term research project aiming to evaluate the full HVP diagram, by including the isospin-breaking effects as well as the disconnected terms. The novelty of our approach is the use of C$^{\star}$ boundary conditions, which allows for defining QED on the lattice with a local and gauge-invariant formulation. The configurations used for this work have been generated by the RC$^{\star}$ collaboration using the \texttt{openQ*D-1.1} code \cite{Campos:2019kgw}. The lattice setup and the methods for the observable are described in sections \ref{setup} and \ref{methods}. Our preliminary results are presented in section \ref{results}.

\section{Lattice setup} \label{setup}

We perform measurements on two QCD ensembles generated by the RC$^{\star}$ collaboration. The configurations are produced at the SU(3) symmetric point, i.e $m_u=m_d=m_s \simeq (m_u^{phys}+m_d^{phys}+m_s^{phys})/3$, by using the Lüscher-Weisz action for the SU(3) field and $O(a)$ improved Wilson fermions. The ensembles are generated with periodic boundary conditions in time and C$^{\star}$ boundary conditions in the spatial directions, i.e. all the fields are periodic up to charge conjugation
\begin{align}\label{eq: Cstar condition}
    U_\mu(x+L_k \hat{k}) = U_\mu^*(x), \qquad
    \psi_f(x+L_k\hat{k}) = C^{-1} \overline{\psi}_f^T(x), \qquad
    \overline{\psi}_f(x+L_k\hat{k}) = -\psi_f^T(x) C.
\end{align}
The action parameters, lattice sizes, and pion masses are shown in Table~\ref{tab: ensembledetails}. 
\begin{table}[h!]
    \centering
    \scalebox{0.9}{%
    \begin{tabular}{c| c c c c c c c c}
    \hline
         Ensemble & V & $\beta$ & $\kappa_{u,d,s}$ & $\kappa_c$& $c_{\text{sw,SU(3)}}$ & $a$ [fm] & $m_{\pi^{\pm}}$ [MeV]    \\
         \hline
         A400a00b324 & $64 \times 32^3$& 3.24  & 0.1344073 & 0.12784 & 2.18859 & 0.05393(24) & 398.5(4.7)  \\
         B400a00b324 & $80 \times 48^3$& 3.24 & 0.1344073 & 0.12784 & 2.18859& 0.05400(14) & 401.9(1.4) \\
         \hline
    \end{tabular}
    }
    \caption{Parameters of the ensembles used in this work. The lattice spacings and pion masses have been computed in Ref.~\cite{https://doi.org/10.48550/arxiv.2209.13183}.}
    \label{tab: ensembledetails}
\end{table}
More details about the tuning of the parameters in the simulations, the scale setting, and the calculations of the meson masses are given in Ref.~\cite{ https://doi.org/10.48550/arxiv.2209.13183}.
In particular, the values of the lattice spacing in Table~\ref{tab: ensembledetails} are determined from the auxiliary scale $t_0$ with the reference value of the CLS determination $(8t_0)^{1/2}=0.415$~fm \cite{Bruno17}.
The two ensembles are generated with the same bare parameters but different lattice volumes. This gives us the possibility to get an idea about the finite-volume effects.
To obtain the results shown in section \ref{results} we use respectively 200 and 108 independent configurations for the ensembles A400a00b324 and B400a00b324.

%Thus, the lattice spacing is not evaluated yet by matching the experimental and lattice value of a physical observable.
%\javadcomment{Let us discuss about this sentence.}
%The current uncertainty on $a$ is only statistical, coming from the propagation of the error on $t_0$. 

\section{Methods for the hadronic vacuum polarization} \label{methods}

In the time-momentum representation (TMR) \cite{Bernecker_2011}, the leading HVP contribution to $a_{\mu}=(g_{\mu}-2)/2$ is given  by the convolution
\begin{equation}
    a_{\mu}^{\mathrm{HVP}} = \left( \frac{\alpha}{\pi} \right)^2 \sum_{t=0}^{\infty} G(t) \tilde{K}(t; m_{\mu}) \label{eq:convolution},
\end{equation}
where $G(t)$ is the spatially summed correlator of two electromagnetic currents
\begin{align}\label{eq: correlator}
    G(t)=- \frac{1}{3} \sum_{k=1,2,3} \sum_{\vec{x}}\ev{V_{k}(x) V_{k}(0)},
\end{align}
and $\tilde{K}(t; m_{\mu})$ is the QED kernel, for which we use the expression in Appendix B in \cite{Morte_2017}. 
There are two commonly used discretizations of the vector current in lattice QCD: the local vector current
\begin{equation}\label{eq: local}
    V^{l}_{\mu,f}(x)=\bar{\psi}_f(x) \gamma_{\mu} \psi_f(x),
\end{equation}
and the point-split or conserved one defined by
\begin{equation} \label{eq: point-split}
        V_{\mu,f}^{c}(x) = \frac{1}{2} \Big[ \bar{\psi}_f(x+\hat{\mu}) \left( 1 + \gamma_\mu \right) U_{\mu}^{\dagger}(x) \psi_f(x) - \bar{\psi}_f(x) \left( 1 - \gamma_\mu \right) U_{\mu}(x) \psi_f(x+\hat{\mu}) \Big],
\end{equation}
where we use the label $f$ to denote the vector current operator of a single flavor.
By inserting the expression of the current in the expectation value in equation \eqref{eq: correlator} and considering  all the possible Wick contractions between the fields, one obtains two different types of contributions:
the connected terms that are flavor diagonal, and the disconnected diagonal and off-diagonal $(f' \neq f)$ terms,
\vspace{8pt}
\begin{equation}
\begin{fmffile}{feyngraph}
      \ev{V_{k}(x) V_{k}(0)}
    = \sum_{f} q_f^2 \times \quad\parbox{25pt}{
    \begin{fmfgraph*}(22,22)
       \fmfleft{v1}
       \fmfdot{v1}
        \fmfv{decor.shape=circle,decor.filled=full,decor.size=4}{v1}
       \fmfright{v2}
       \fmfdot{v2}
        \fmfv{decor.shape=circle,decor.filled=full,decor.size=4}{v2}
       \fmf{fermion,left,tension=0.1,label=$f$}{v1,v2,v1}
    \end{fmfgraph*}}
    \quad
    + \sum_{f, f'} q_{f}q_{f'} \times \quad\parbox{25pt}{
    \begin{fmfgraph*}(22,22)
    \fmfleft{v3}
       \fmfdot{v3}
        \fmfv{decor.shape=circle,decor.filled=full,decor.size=4}{v3}
       \fmfright{v4}
       \fmf{fermion,left,tension=0.1,label=$f$}{v3,v4,v3}
       
       \fmfleft{v5}
       \fmfdot{v5}
        \fmfv{decor.shape=circle,decor.filled=full,decor.size=4}{v5}
       \fmfright{v6}
       \fmf{fermion,left,tension=0.1,label=$f$}{v6,v5,v6}
    \end{fmfgraph*}}
    \quad\parbox{25pt}{
    \begin{fmfgraph*}(22,22)
    \fmfleft{v5}
       \fmfright{v6}
       \fmfdot{v6}
        \fmfv{decor.shape=circle,decor.filled=full,decor.size=4}{v6}
       \fmf{fermion,right,tension=0.1,label=$f'$}{v6,v5,v6}
       \end{fmfgraph*}} \,.
\end{fmffile}
\end{equation}
\vspace{2pt}

\noindent In the following, we will focus only on the connected terms.

The local vector current in equation \eqref{eq: local} is neither conserved nor improved on the lattice. If we consider only the connected contractions, it renormalizes independently for each flavor $f$
\cite{Bhattacharya_2006,G_rardin_2019}
\begin{equation} \label{eq: renormalized}
    V^{R}_{\mu,f}=Z_V^{m_f} (V_{\mu,f}^{l} +a c_V \partial_{\nu} T_{\mu\nu,f}),
\end{equation}
where $T_{\mu \nu, f}=-\bar{\psi}_f \frac{1}{2}[\gamma_{\mu}, \gamma_{\nu}] \psi_f$ is the tensor current, $c_V$ is a constant, and $m_f$ is the mass of the valence quark with flavor $f$. The current in equation \eqref{eq: point-split} is instead conserved on the lattice but still requires $O(a)$ improvements. In this work, we do not consider any improvements at the observable level, thus we neglect the term proportional to the tensor current. In this case, we see from equation \eqref{eq: renormalized} that the local vector current for a flavor $f$ renormalizes multiplicatively through the mass-dependent renormalization factor $Z_V^{m_f}$. We describe our method to determine $Z_V^{m_f}$ in section \ref{sec: renormalization }. The choice of the local or conserved currents at the source and sink points of the quark propagator leads to different discretizations of the correlator $G(t)$ in TMR, but share the same continuum limit once renormalization constants are taken into account. 

\subsection{Signal-to-noise ratio}

Before performing the measurements, we study the effect of the discretization of the current on the signal-to-noise ratio of the correlator. By using the two expressions of the current in equations \eqref{eq: local} and \eqref{eq: point-split}, it is indeed possible to define three types of correlator: the local-local ($ll$), the conserved-conserved ($cc$), and the mixed one ($cl$).
For instance, for the local-local correlator, the expression to be evaluated is the following
\begin{align}\label{eq:em_traces}
    G^{ll}_{f}(t)_{conn} =
    & \frac{1}{3}\sum_{k=1,2,3}\sum_{\vec{x}} q_f^2 \tr \left[ \gamma_k D_f^{-1}(x|0) \gamma_k D_f^{-1}(0|x) \right], 
\end{align}
with $D^{-1}(x|0)$ being the quark propagator from $0$ to $x$. For these measurements, we use 60 configurations and 10 point sources per configuration. The aim is to understand which choice is the most convenient in terms of signal-to-noise ratio and computational cost.
With the conserved-conserved correlator, we do not need to determine the renormalization factor.
However, we expect to have a noisier result when using the conserved current due to the fluctuations of the gauge field. Moreover, employing the conserved current both at the sink and source points requires 3 additional inversions of the Dirac operator per point source, one for each spatial direction $\hat{k}=1,2,3$.
\begin{figure}
\begin{minipage}{0.5\textwidth}
        \centering
        \includegraphics[width=1.0\textwidth]{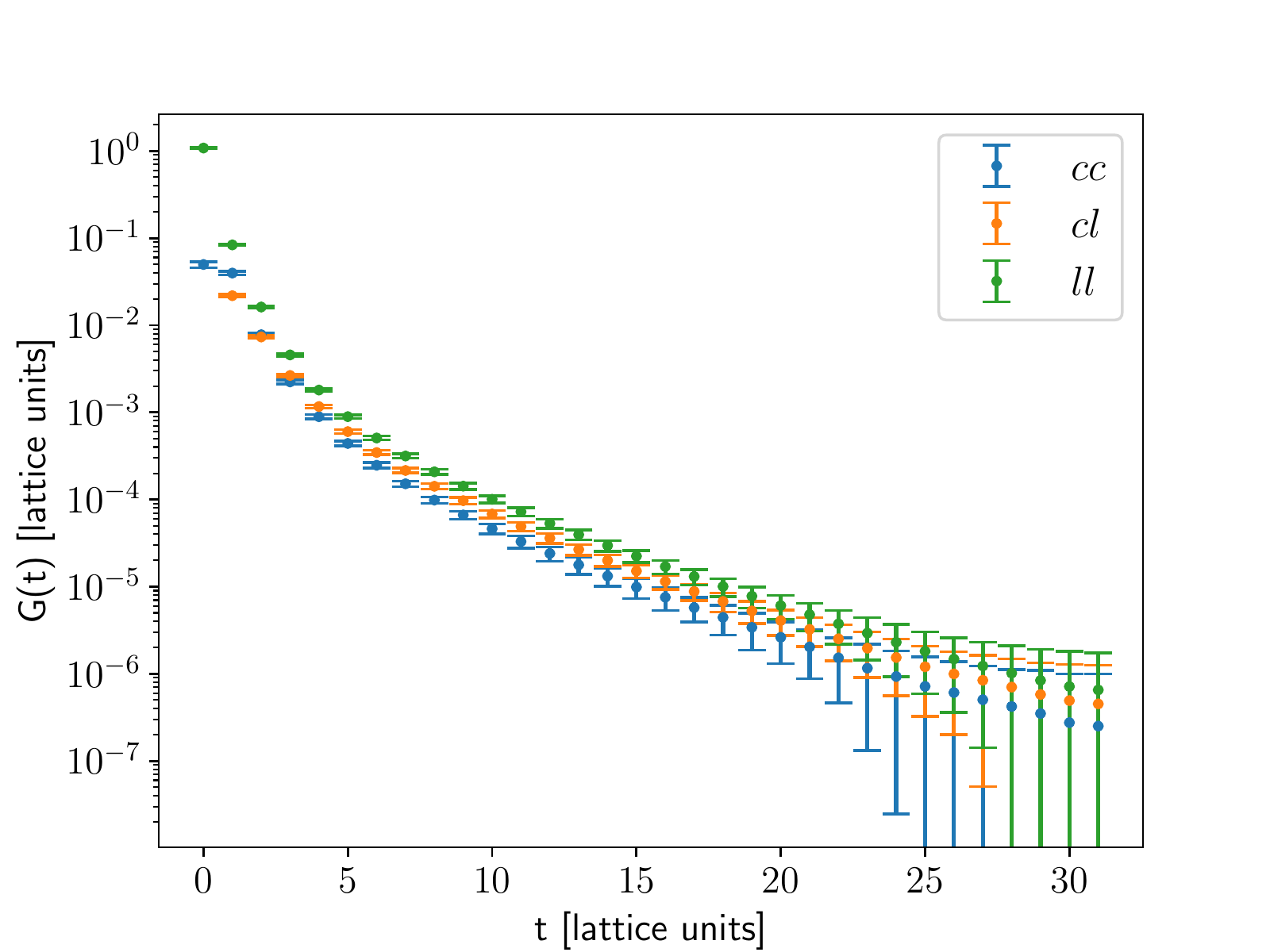} % first figure itself
      
    \end{minipage}\hfill
    \begin{minipage}{0.5\textwidth}
        \centering
        \includegraphics[width=1.0\textwidth]{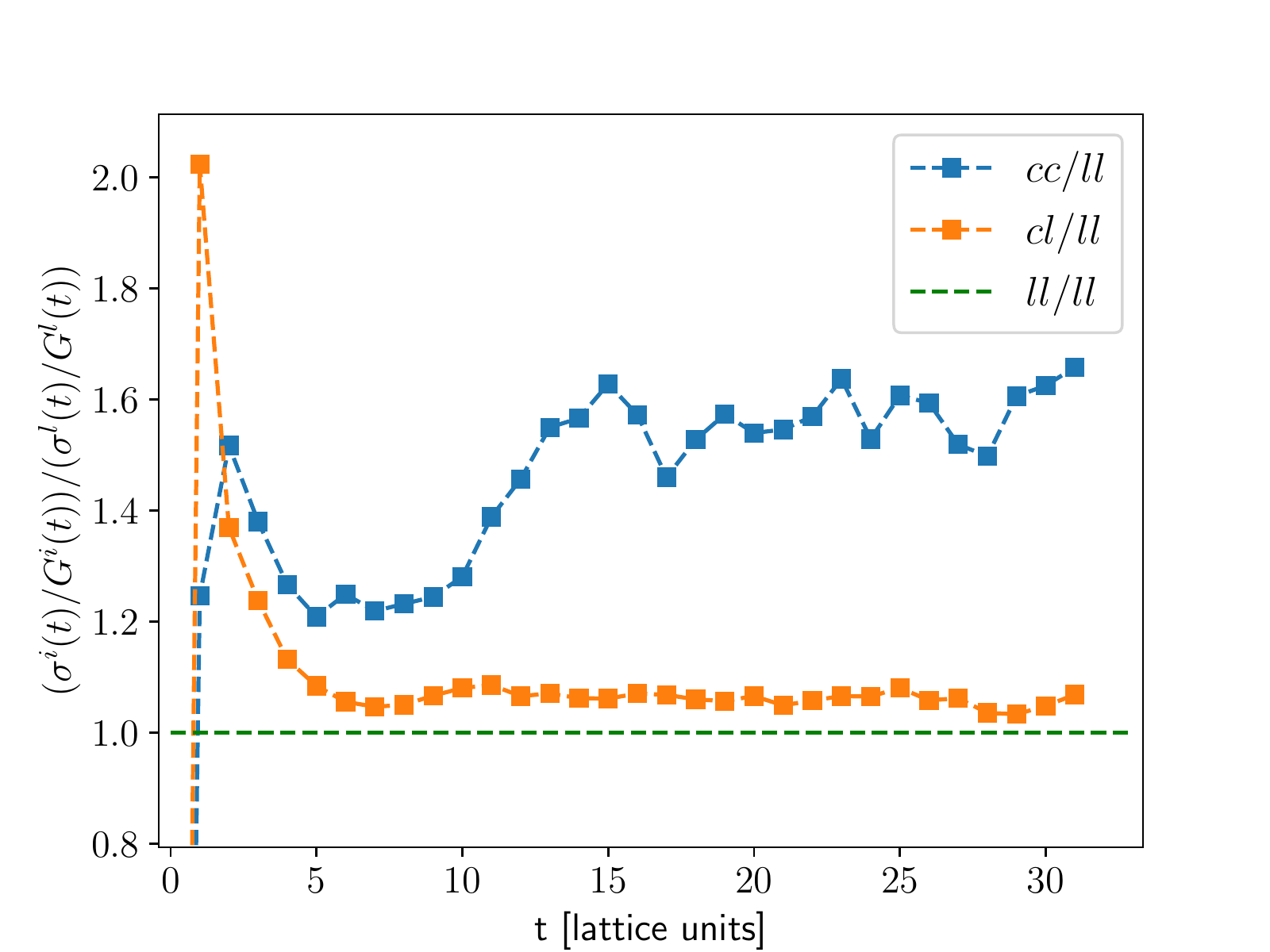} % second figure itself
        
    \end{minipage}
      \caption{Example of comparison of the correlators $G^{kk}(t)$, with $k=c,l$ (left), and the relative errors (right) for the light quark. $c$ and $l$ denote the conserved and the local discretization of the vector current.}
      \label{fig: three corr}
\end{figure}

Figure \ref{fig: three corr} shows the three different correlators of the light quark measured on the A400a00b324 ensemble. The left panel shows the correlators plotted against time, and the right panel illustrates the relative statistical noise of $G^{cl}(t)$ and $G^{cc}(t)$ compared to the local-local correlator. As shown, the conserved-local correlator is only slightly (5\% to 10\%)
noisier than the local-local one; the conserved-conserved correlator is instead much noisier. By taking into account that the computational cost is even four times larger, the conserved-conserved correlator is not a good choice to achieve the overall target precision.
The other two correlators are equivalent choices unless the uncertainty in $Z_V$ gets significant, then the conserved-local correlator has the advantage to be less sensitive to the precision of the renormalization factor since it appears only once in this correlator.

In section \ref{results} we will show results for both local-local and local-conserved correlators, pointing out the significant difference in the charm contribution, due to large discretization effects.

\section{Strange and charm quark-connected contribution} \label{results}

\subsection{Tuning procedure}\label{sec: tuning}
To evaluate the leading order strange and charm quark-connected contribution to HVP, it is necessary to perform the continuum limit and the extrapolation to the physical pion mass and take into account all the systematics. In this work, we consider only one value for the lattice spacing and pion mass and two different volumes. Before evaluating the correlator in equation \eqref{eq:em_traces}, we tune the hopping parameters $\kappa_f$ of the valence quarks.
We choose the value of $\kappa_{s}$ and $\kappa_{c}$ by matching the physical value of the masses of the mesons $\phi$ and $J\slash \psi$ \cite{Workman:2022ynf}
\begin{equation}\label{eq: phys masses}
    m^{phys}_{\phi} =1019.461(20)\; \text{MeV},
    \qquad  m^{phys}_{J \slash \psi} = 3096.900(6) \; \text{MeV}
\end{equation}
with our lattice results, obtained respectively from the two-point functions of the interpolators
%We choose the value of $\kappa_{s}$ and $\kappa_{c}$ based on the vector mesons interpolated by the operators
\begin{align}
    \mathcal{O}_s=\bar{s}\gamma_{\mu}s, \qquad
    \mathcal{O}_c=\bar{c}\gamma_{\mu}c.
\end{align}
In this matching procedure we are neglecting the disconnected terms and the QED corrections, which enter into the physical masses and are instead missing in our calculations.
%We require that the effective masses obtained from the corresponding two-point functions match the physical values \cite{Workman:2022ynf}

%Another possible strategy for tuning the hopping parameters is to use the pseudoscalar mesons $\eta_{s,c}$. The two strategies should lead to the same results for $a^{HVP,s/c}_{\mu}$ when considering the chiral-continuum limit. Thus, both of them can be included in the analysis and used for estimating the final systematic error. 
In Tables \ref{tab: small ensemble} and \ref{tab: large ensemble}  we show the different choices of $\kappa_{s/c}$ and the results for the effective masses of the vector mesons $s\bar{s}$ and $c\bar{c}$ for both ensembles.
\begin{table}[h!]
        \centering
        \scalebox{0.9}{%
        \begin{tabular}{c c c | c c c}
        \hline
          $\kappa_s$&  $am_V(s\bar{s})$ & $m_V(s \bar{s})$ [MeV] & $\kappa_c$ & $am_V(c\bar{c})$ & $m_V(c \bar{c})$ [MeV]  \\ [0.5ex] 
          \hline
          0.134407 & 0.2644(50) & 967(19) & 0.12784 & 0.8540(5) & 3125(14)  \\  
         0.1343   & 0.2731(24) & 999(10) & 0.12794 & 0.8463(5) & 3097(14) \\ 
         0.13422   & 0.2808(22) & 1027(9) & 0.12800 & 0.8418(5) & 3080(14) \\ 
         \hline

        \end{tabular}
        }
        \vspace{10pt}
        \caption{Ensemble A400a00b324: mass of the vector mesons for several choices of the hopping parameters in the valence sector. Values in MeV are obtained by using the reference value $(8t_0)^{1/2}=0.415$ fm \cite{https://doi.org/10.48550/arxiv.2209.13183}.}
        \label{tab: small ensemble}
\end{table}

\begin{table}[h!]
        \centering
        \scalebox{0.9}{%
        \begin{tabular}{c c c| c c c}
        \hline
          $\kappa_s$&   $am_V(s\bar{s})$ & $m_V(s \bar{s})$ [MeV] & $\kappa_c$ & $am_V(c \bar{c})$ & $m_V(c\bar{c})$ [MeV]  \\ [0.5ex] 
         \hline
          0.134407 & 0.2522(33)  & 923(13) & 0.12784 & 0.8536(7) & 3123(14) \\
         0.134220 & 0.2715(22)  &  993(9) & 0.12794 &  0.8458(9) &3095(14)\\ 
         0.134152   & 0.2794(19) & 1022(8)   \\ 
        \hline
        \end{tabular}
        }
        \vspace{10pt}
        \caption{Ensemble B400a00b324: mass of the vector mesons for several choices of the hopping parameters in the valence sector. Values in MeV are obtained by using the reference value $(8t_0)^{1/2}=0.415$ fm \cite{https://doi.org/10.48550/arxiv.2209.13183}.}
        \label{tab: large ensemble}
\end{table}

\begin{figure}
\begin{minipage}{0.5\textwidth}
    \centering
    \includegraphics[width=1.05 \linewidth]{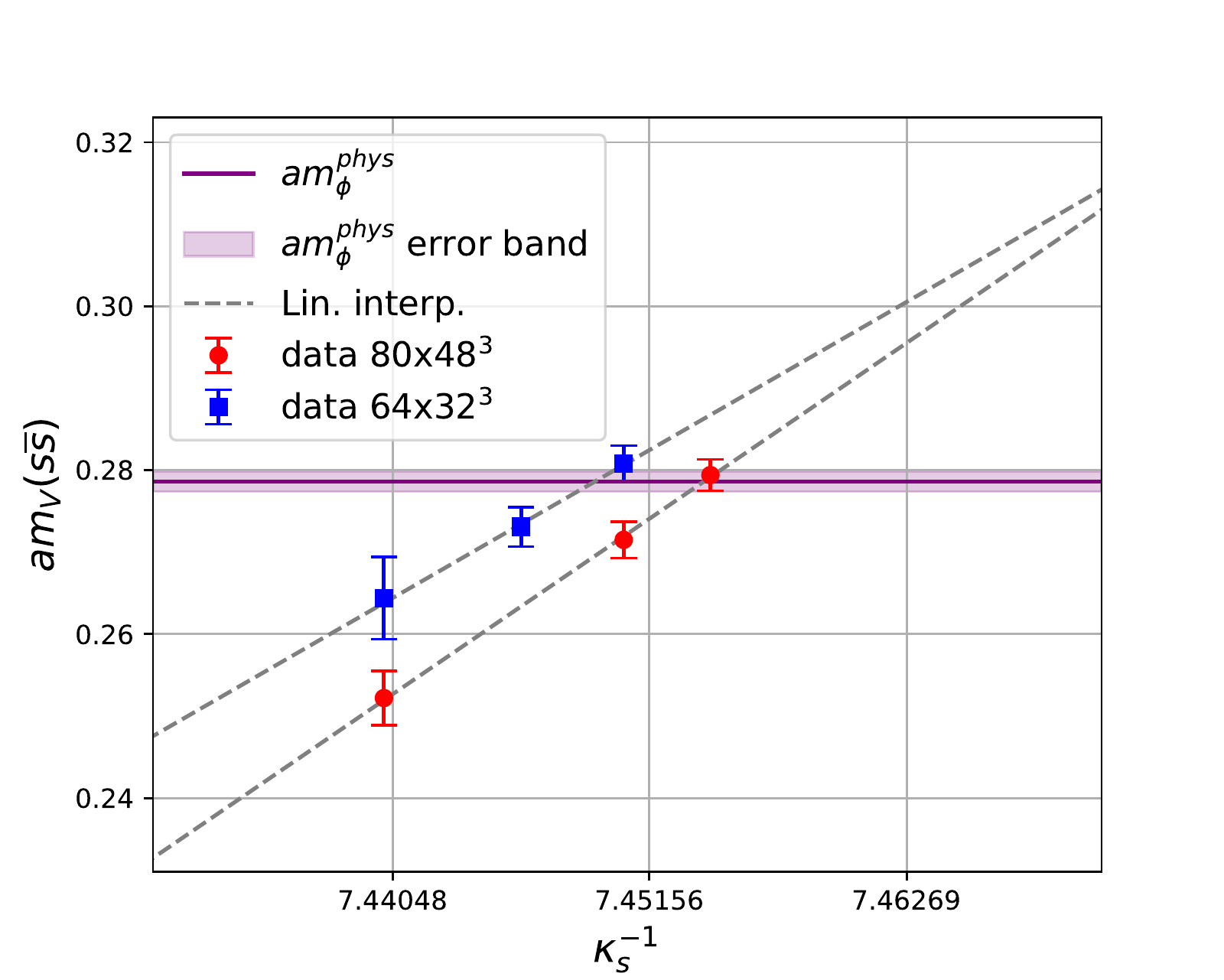}
\end{minipage}
\begin{minipage}{0.5\textwidth}
    \centering
    \includegraphics[width=1.052 \linewidth]{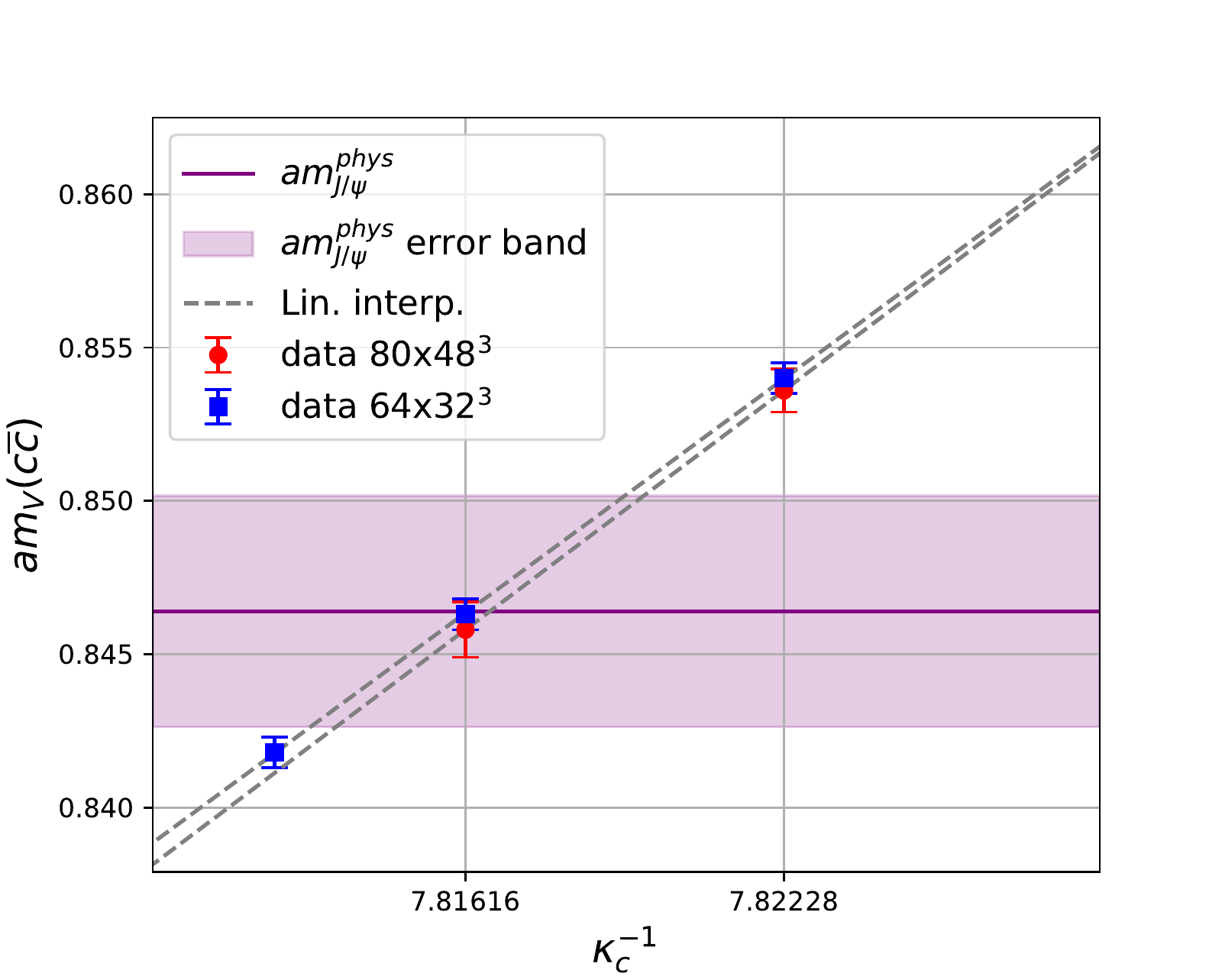}
    
    \end{minipage}
    \caption{Masses of the vector mesons $s\bar{s}$ (left) and $c \bar{c}$ (right) as functions of the inverse of the hopping parameter. The purple bands and their central value represent the physical mass of the mesons $\phi$ (left) and $J \slash \psi$ (right) converted to lattice units.}
     \label{fig: tuninglocloc}
\end{figure}

We plot the masses of the vector mesons as a function of the inverse of the corresponding hopping parameters $\kappa^{-1}_{s/c}$, which are linear in the bare masses of the valence quarks $s$ and $ c$. 
In Fig. \ref{fig: tuninglocloc} we show this dependence for both strange (left) and charm (right) quarks. The purple bands in the plots correspond to the physical masses in equation \eqref{eq: phys masses} converted to lattice units. %The errors come essentially from the propagation of the error on $a$. From the figure, it is clear that a very precise tuning of $\kappa_c$ is not required at the current uncertainty of $a$, which is the dominant source of error in this case. The situation for the vector meson $\phi$ is different: the statistical error of $am_{\phi}$ and the uncertainty on $a$ equally contribute to the precision of the tuning procedure. 

\subsection{Renormalization constants} \label{sec: renormalization }
\begin{figure}[h!]
    \centering
    \includegraphics[width=0.55 \linewidth]{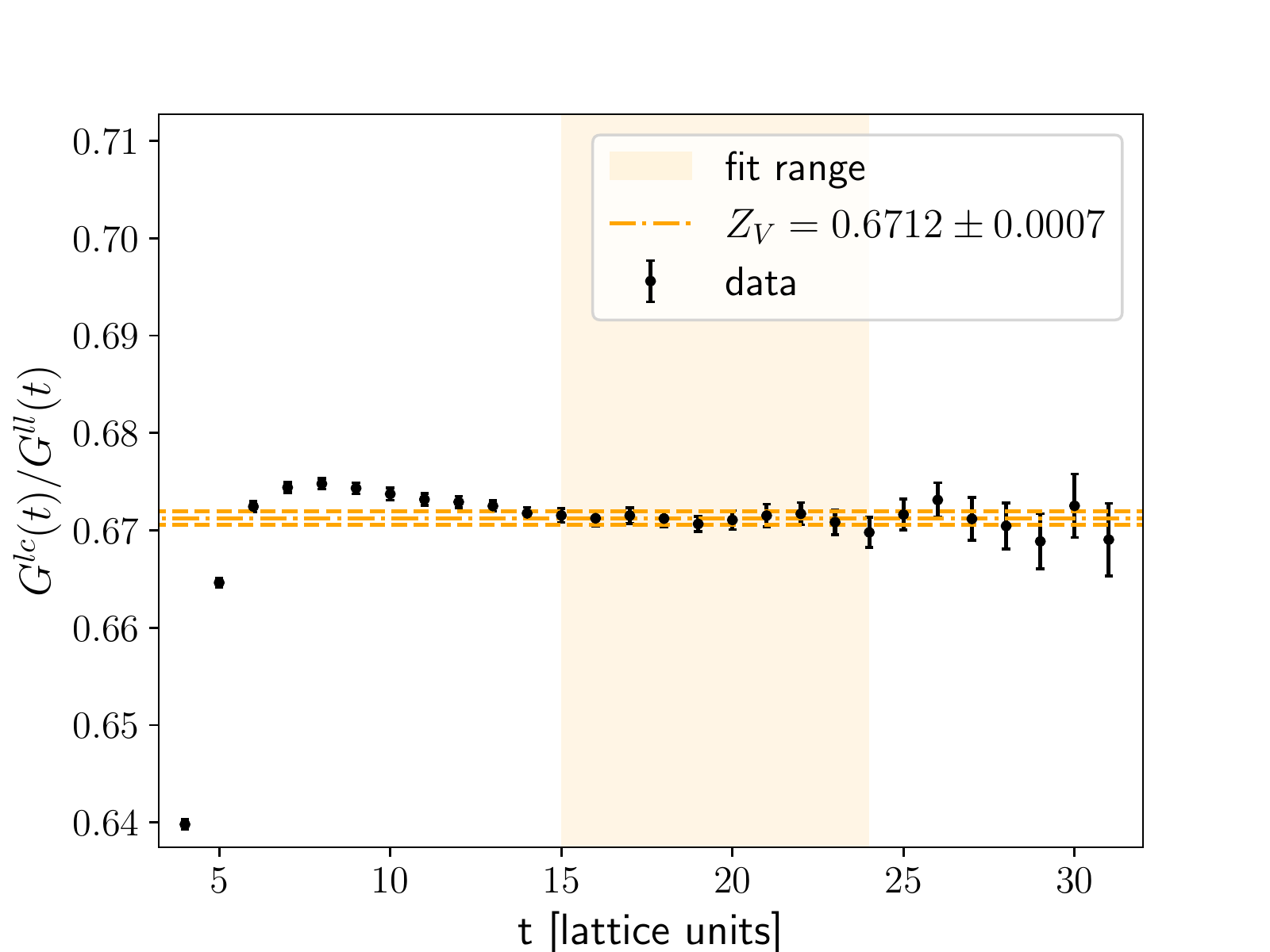}
    \caption{Determination of the renormalization constant $Z_V^{m_s}$ for the local vector current with $\kappa^{tun}_{s}=0.13422$ on the A400a00b324 ensemble.}
    \label{fig: Zv light A400}
\end{figure}

Evaluating $a^{HVP, s/c}_{\mu}$ from the local-local or conserved-local correlators requires determining the renormalization factor $Z_V^{m_f}$ and the improvement coefficient introduced in \eqref{eq: renormalized}. In this work, we do not consider any improvement terms and evaluate the mass-dependent renormalization factor $Z_V^{m_f}$ of the local current from the ratio \cite{Boyle_2017}
\begin{equation}
    R(t)=\frac{\sum_{\vec{x},k}\ev{V^{c}_{k,f}(x) V^{l}_{k,f}(0)}}{\sum_{\vec{x},k}\ev{V^{l}_{k,f}(x) V^{l}_{k,f}(0)}}\, .
    \label{eq:ratio-method-for-Zv}
\end{equation}
When $t$ is small, the quantity is affected by the different discretization effects of the two currents. At large time, $R(t)$ saturates and we can determine $Z_V^{m_f}$ by fitting the plateau region to a constant. An example of such a fit is shown in Fig. \ref{fig: Zv light A400}. We applied the same method for both ensembles and for both $Z_V^{m_s}$ and $Z_V^{m_c}$.

In Table \ref{tab: Zv}  we show the fit ranges and the values obtained for $Z_V^{m_{s/c}}$ for the tuned hopping parameters $\kappa^{tun}_{s/c}$.
\begin{table}[h!]
    \centering
    \scalebox{0.9}{%
    \begin{tabular}{c|c c | c c }
    \hline
         Ensemble & fit range & $Z^{m_s}_V$ & fit range & $Z^{m_c}_V$\\
         \hline
         A400a00b324& [15,24] & 0.6712(7) & [24,30] & 0.6066(2) \\
         B400a00b324& [15,24] & 0.6707(5) & [23,31] & 0.6066(4) \\
         \hline
    \end{tabular}
    }
    \caption{Mass-dependent renormalization factor obtained from the ratio method
    defined in Eq.~\eqref{eq:ratio-method-for-Zv}.}
    \label{tab: Zv}
\end{table}
The errors are determined with the bootstrap procedure.
\subsection{Results}

The evaluation of the leading HVP contribution to $a_{\mu}$ requires an integration over the euclidean time 
\begin{equation}\label{eq: integral in x0}
    a_{\mu}^{\mathrm{HVP},f} = \left( \frac{\alpha}{\pi} \right)^2 \sum_{t=0}^{\infty} G^{f}(t) \tilde{K}(t; m_{\mu}).
\end{equation}
One of the problems related to this task is that the signal deteriorates with the lattice time $t$, due to the exponentially increasing errors of the correlator. Another related issue comes from the finite size of the box: the integration domain is indeed restricted to $[0,T/2]$, then we have to extrapolate the correlator to infinite time.
In addition, the correlator is affected by finite-volume effects (FVE) due to the finite temporal ($T$) and spatial ($L$) extents.

Concerning the finite-volume effects, it has been found \cite{Hansen_2020} that for given $L$ the leading finite-$L$ corrections are the exponentials $e^{-m_{\pi}L}, e^{-m_{\pi} \sqrt{2}L}$ and $e^{-m_{\pi}\sqrt{3} L}$. Similarly, the leading contribution arising from finite $T$ is $e^{-m_{\pi}T}$.
As a consequence, the finite-$T$ effects are higher order corrections since usually in the simulations $T=2L$. 
These results have been derived for a periodic torus in four dimensions and are affected by the choice of the boundary conditions. In our setup, the boundary condition in the time direction is periodic, then the results for finite-$T$ corrections found in
Ref.~\cite{Hansen_2020} still apply. However, we use C$^{\star}$ boundary conditions in all three spatial directions, which means that the finite-$L$ corrections are in general different. Some studies have shown that in pure QCD C$^{\star}$ boundary conditions lead to small improvements for the FVE, with a leading correction $e^{-m_{\pi}\sqrt{2}L}$ \cite{Sophietalk}.
A detailed numerical study of the finite-volume effects for ensembles with C$^{\star}$
boundary conditions will be  carried out in future work. In this work, we make a direct comparison of the results for the integrand $G(t)\tilde{K}(t,m_{\mu})$ and $a^{\mathrm{LO,HVP}}_{\mu}$ on the two available QCD ensembles.

To control the large-time behavior of the correlator, we use the following quantity
\begin{align}
    G_\text{constructed}(t) = 
    \begin{cases}
       G(t) \qquad &(t \le t_{0, \text{cut}})\\
       G_{\text{1-exp}}(t) \qquad  &(t_0 > t_{0, \text{cut}})
    \end{cases}
\end{align}
where $t_{0,  \text{cut}}$ is a properly chosen cut-off and $G_{\text{1-exp}}(t)$ denotes the exponential extrapolation of the correlator at large time
\begin{equation}
    A \exp(-t \cdot m_\text{eff}).
\end{equation}
The two parameters $m_\text{eff}$ and $A$ are the effective mass and the amplitude obtained through a fit procedure to the correlator. For the masses, we use the results reported in Tables \ref{tab: small ensemble} and \ref{tab: large ensemble} for the tuned hopping parameters.
The parametrization with a single-exponential is a crude approximation that introduces some systematics since we are neglecting the excited states contributing to the correlator. We plan to use a more accurate model for the tail of the correlator in future works.
\begin{figure}
\begin{minipage}{0.5\textwidth}
    \centering
    \includegraphics[width=1.0 \linewidth]{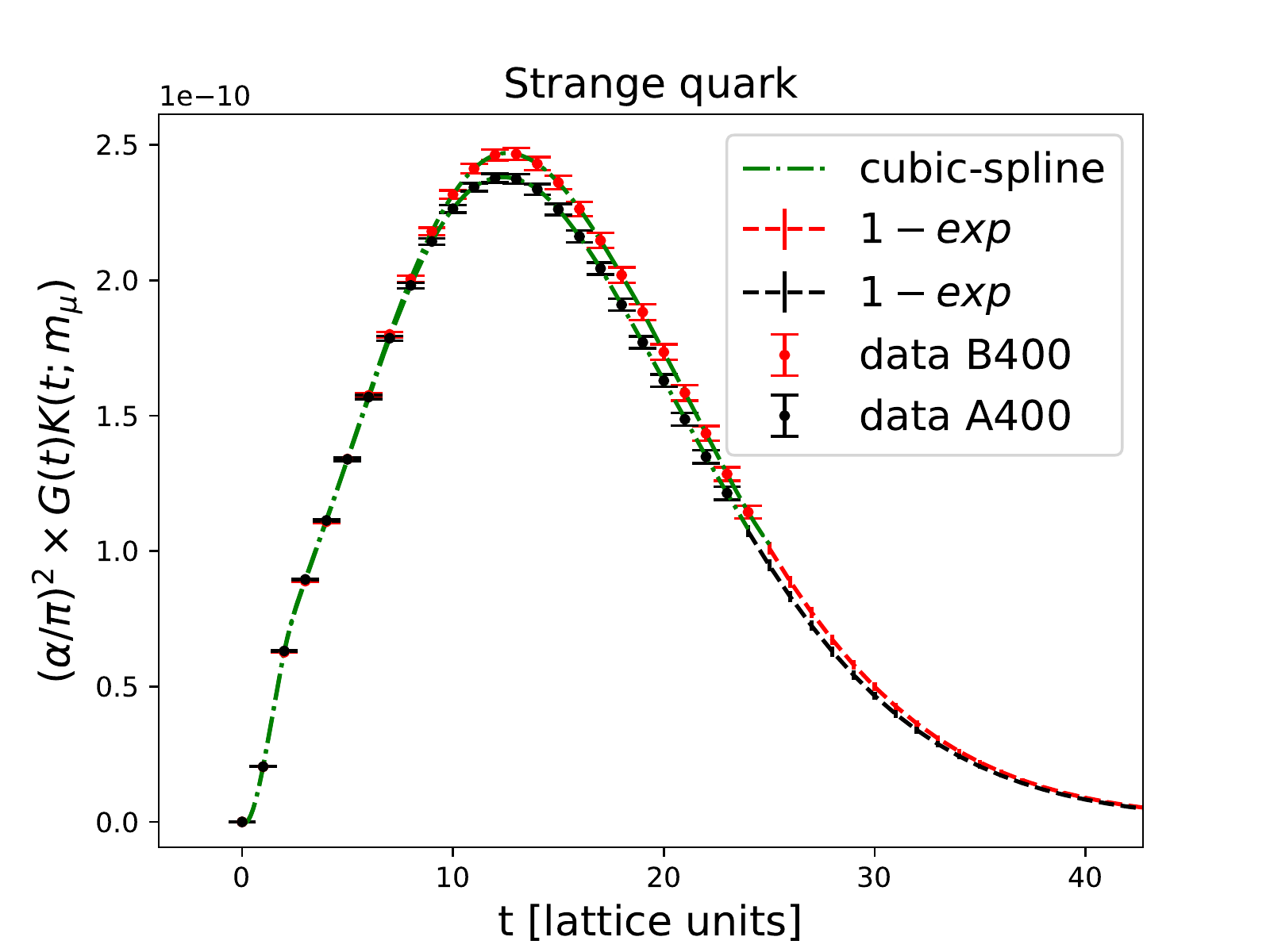}
\end{minipage}
\begin{minipage}{0.5\textwidth}
    \centering
    \includegraphics[width=1.0 \linewidth]{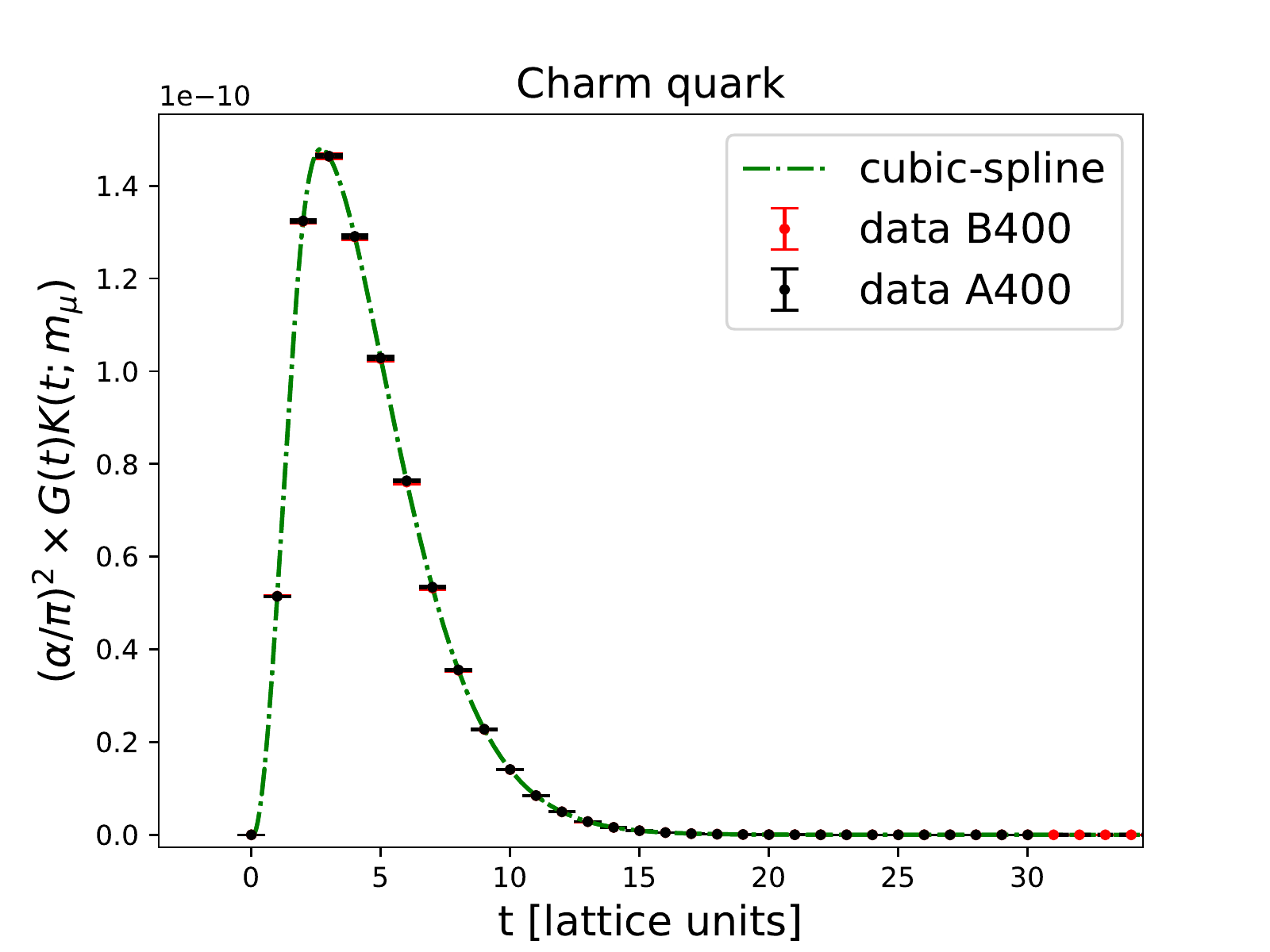}
\end{minipage}
    \caption{Comparison of the integrands for the strange (left) and charm (right) contributions between the two ensembles. The tail of the integrands for the strange contribution is approximated by a single exponential.}
    \label{fig: plot integrand}
\end{figure}

The plots in Fig. \ref{fig: plot integrand} show the integrands both for the charm and strange quarks contributions and the two ensembles. The lattice data for the charm contribution are sufficiently precise and do not require any extrapolation or improvement. In the case of the strange contributions, the tail of the integrand is approximated as described above.
The results of the integration are listed in Table \ref{tab: summary results}. We estimate $a^{\mathrm{LO,HVP}}_{\mu}$ using the two different discretizations of the correlator: conserved-local and local-local. The strange contribution is not affected by the choice of the correlator, the results are indeed compatible with the current uncertainties for both ensembles. By contrast, the contribution from the charm quark is particularly sensitive to the choice of discretization.
The finite-size effects are negligible for the charm quark contribution and lead instead to a difference of about $2\sigma$ for the strange quark.
\begin{table}[h!]
    \centering
    \scalebox{0.9}{% 
    \begin{tabular}{c|c|c|c}
    \hline
         Ensemble & Type & $a^{s}_{\mu} \times 10^{-10}$  &
         $a^{c}_{\mu} \times 10^{-10}$  \\
         \hline
          \multirow{2}{*}{A400a00b324} & $ll$  & 46.7(7) & 7.83(8)\\
          & $cl$  &  46.2(7) & 6.18(7) \\
         \hline
         \multirow{2}{*}{B400a00b324} &  $ll$ & 48.5(7) & 7.81(9) \\
          & $cl$ & 48.0(7) & 6.16(7) \\
         \hline
    \end{tabular}
    }
    \caption{ Results for $a^{s,c}_{\mu}$ in  units of $10^{-10}$ determined using the TMR and two different discretizations of the observable: local-local and conserved-local.}
    \label{tab: summary results}
\end{table}
\subsection{Partial errors budget}
The errors in Table \ref{tab: summary results} are the quadratic sum of the statistical and part of the systematic errors as follows. The uncertainties taken into account are the statistical errors from the correlators, the lattice spacing and $Z_V$, and the systematics from the choices of the cut-off (for the strange quark) and the fit range used to determine $A$ and $m_\text{eff}$.
The statistical error is determined by using the bootstrap method. Since $Z_V$ appears as a multiplicative factor in front of the whole expression, we employ the standard error propagation for it.
The lattice spacing's values are determined by using the reference scale $(8t_0)^{1/2}=0.415$ fm as an absolute value, without taking into account the systematics coming from the uncertainty on this scale. Thus, our current error on $a$ is only a statistical partial uncertainty. The dependence on the lattice spacing is in the QED kernel $\tilde{K}(t, m_{\mu})$: we numerically propagate the partial error $\delta a$ by repeating the evaluation of $a_{\mu}$ for $N$ values of the lattice spacing drawn from a normal distribution $\mathcal{N}(a, \delta a)$. We use at least $N=100$ for each result. Finally, we repeat the calculation for several values of the fit range and the cut-off and apply a weighted averaging procedure to get the total systematics.

%The pie charts in Fig. \ref{fig: error} represent the relative contribution of the accounted uncertainties to the final error. The current precision is about 1.5\% and 1.1\% for $a^s_{\mu}$ and $a^c_{\mu}$ respectively.  For the strange contributions, the statistical error and the current lattice spacing's precision equally contribute. The lattice spacing is instead the dominant error source for the $a^c_{\mu}$, contributing up to the 90\% to the final precision. 
We remark that there are still several unaccounted uncertainties. We are currently missing the systematics introduced by the single-exponential extrapolation of the correlator and by the use of the reference scale $t_0$ without an error for the determination of the lattice spacing. In this work we did not perform a quantitative numerical study of the finite-size effects and we measured at one value of the lattice spacing and pion mass, thus we have not performed yet an extrapolation of the results to the continuum and physical point.
%\begin{figure}[h!]
 %   \centering
 %   \includegraphics[width=0.6 \linewidth]{img/plot_chart_combined.pdf}
 %   \caption{Partial error budgets for the strange and charm quark contributions. We consider the statistical error of the lattice spacing, the statistical uncertainty of the correlator, and the precision of the renormalization constant. The partial systematic errors included in the left pie chart are the ones due to the choice of the fit range and the cut-off for the single-exponential model used for the tail of the correlator.}
 %   \label{fig: error}
%\end{figure}
 
\section{Conclusions and outlooks}
We have measured the connected contribution to the leading hadronic vacuum polarization from strange and charm quarks, in a setup with C$^{\star}$ boundary conditions in the three spatial directions. We performed the analysis on two ensembles with different volumes, indicating that the finite-size effects are under control. As expected, we find that the charm contribution is considerably affected by the choice of the correlator, due to the sensitivity to the discretization effects. In addition, we have shown that a more precise determination of the lattice spacing is needed to reach the target precision.
Our plans for future works include the evaluation of the isospin-breaking effects as well as the disconnected terms, and a quantitative study of the finite-size effects.

\acknowledgments

We acknowledge access to Piz Daint at the Swiss National Supercomputing Centre, Switzerland under the ETHZ's share with the project IDs s1101, eth8 and go22.
Financial support by the SNSF (Project No. 200021\textunderscore200866) is gratefully
acknowledged.
L.B., S.M., and M.K.M received funding from the European Union's Horizon 2020 research and innovation programme under the Marie Sk\l odowska-Curie grant agreement No. 813942. M.D.received funding from the European Union’s Horizon 2020 research and innovation programme under grant agreement No. 765048.
A.C.’s and J.L.’s research is funded by the Deutsche
Forschungsgemeinschaft Project No. 417533893/ GRK-2575 “Rethinking Quantum Field Theory”.

\bibliographystyle{JHEP} %siam(ordine alf.) ieeetr (apparizione)
\bibliography{bibliography.bib}
%\begin{thebibliography}{99}
%\bibitem{...}
%\end{thebibliography}

\end{document}